# Bas-relief Generation from Point Clouds Based on Normal Space Compression with Real-time Adjustment on CPU

Jianhui Nie, Wenkai Shi, Ye Liu, Hao Gao, Member, IEEE,
Feng Xu, Zhaochen Zhang and Guoping Jiang, Senior Member, IEEE

**Abstract**—Bas-relief generation based on 3d models is a hot topic in computer graphics. State-of-the-art algorithms take a mesh surface as input, but real-time interaction via CPU cannot be realized. In this paper, a bas-relief generation algorithm that takes a scattered point cloud as input is proposed. The algorithm takes normal vectors as the operation object and the variation of the local surface as the compression criterion. By constructing and solving linear equations of bas-relief vertices, the closed-form solution can be obtained. Since there is no need to compute discrete gradients on a point cloud lacking topology information, it is easier to implement and more intuitive than gradient domain methods. The algorithm provides parameters to adjust the bas-relief height, saturation and detail richness. At the same time, through the solution strategy based on the subspace, it realizes the real-time adjustment of the bas-relief effect based on the computing power of a consumer CPU. In addition, an iterative solution to generate a bas-relief model of a specified height is presented to meet specific application requirements. Experiments show that our algorithm provides a unified solution for various types of bas-relief creation and can generate bas-reliefs with good saturation and rich details.

——————————— ◆ ———————————

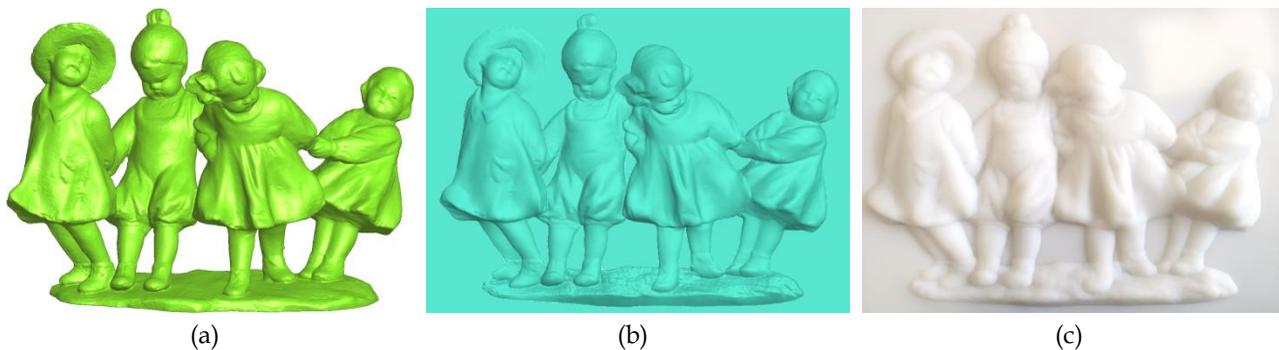

(a)          (b)          (c)

Fig. 1 A 3D scene data containing multiple models (a), the bas-relief model generated by our algorithm (b), and its 3D printing result(c).

## 1 INTRODUCTION

A bas-relief is a form of sculpture that compresses 3d models into a limited space with a plane or curved surface as the carrier. It is widely used in product packaging, appearance decoration, commemorative coin production and other industries [1]~[3]. The traditional bas-relief generation process is complicated and difficult to popularize, while the bas-relief generation method based on 3d models is highly automated and can be completed without the participation of professionals, which has become a hot topic in the field of graphics.

The algorithms for bas-relief generation can be divided into two types: depth domain and gradient domain. The depth domain method directly takes the spatial coordinates of the model as the operating object and optimizes the effect of height compression by means of histogram equalization, nonlinear filtering, etc. Relevant algorithms are simple to implement and fast in operation, but it is difficult to maintain the saturation of the original model when the compression ratio is large. The gradient field method takes the differential coordinates as the operating object and finishes bas-relief reconstruction by maintaining or enhancing the gradient field feature and solving the Poisson equation. Bas-reliefs generated by this type of method can well retain the details of the original model. However, the operation in the gradient field is not intuitive enough, so it is difficult to accurately predict the bas-relief height and effect before the reconstruction is completed.

————————————————
- *J.H. Nie, Y. Liu, W.K. Shi, H. Gao, Z.C. Zhang are with the College of Automation & College of Artificial Intelligence, Nanjing University of Posts and Telecommunications, Nanjing, China. J.H. Nie, Y. Liu and G.P. Jiang are also with the Jiangsu Engineering Laboratory for Internet of Things and Intelligent Robotics, Nanjing, China. E-mail: {njh19, yeliu, jianggp} @ njupt.edu.cn, {wenkaishi.njupt, tsgaohao, zhaochenzhang72} @ gmail.com.*
- *F. Xu is with the school of Software, Tsinghua University, Beijing, China. E-mail: feng-xu@tsinghua.edu.cn.*
*Corresponding author: J.H. Nie*

Currently, most bas-relief generation algorithms take the mesh surface as input, but with the continuous progress of visual measurement technology, especially the widespread application of the Kinect depth sensor, more and more models are represented and rendered in the form of point clouds [4][5]. Meanwhile, as an art activity, real-time interactive adjustment is an essential operation in the creation process, which puts forward high requirements for the execution efficiency of the algorithm, especially in the absence of GPU acceleration. How to achieve real-time adjustment of the bas-relief effect is an urgent problem to be solved. In addition, the current algorithms mostly include parameters that affect the model depth, which can be modified to achieve height adjustment. However, the inverse problem, that is, how to automatically adjust parameters to make the resulting bas-relief conforms to the specified height, has not been studied. In view of the above problems, this paper proposes a bas-relief generation algorithm based on the input of a scattered point cloud. Through normal space compression based on the variation of the local surface, it can realize the real-time adjustment of model height relying only on the computation power of a consumer CPU, and it can also adjust the bas-relief to the specified height according to need. Specifically, the main contributions of this paper are as follows: 1) a bas-relief generation algorithm with a point cloud as input is proposed to realize model height compression through the nonlinear transformation of normal space. The normal vector is the basic attribute of a point cloud. Thus, compared with the gradient domain methods, it is easier and more intuitive to directly operate the normal vector. Meanwhile, the problem can be transformed into solving the linear sparse equations about the height of vertices, and then the closed-form solution can be obtained. 2) We realize the real-time adjustment of the bas-relief effect relying only on the computational power of a consumer CPU. By sampling a sparse control point subset from the visible point cloud and predecomposing the solution matrix, the scale of the problem to be solved is effectively reduced. Meanwhile, the height of the entire point cloud is adjusted though the mapping from subspace to the global space. 3) The algorithm provides parameters, which can flexibly adjust the bas-relief height, saturation and detail richness. Meanwhile, it also provides an iterative solution strategy to generate the bas-relief model with the specified height to better meet users' creation needs.

The following sections are organized as follows. Section 2 reviews relevant work. Section 3 introduces the algorithm of bas-relief generation and its implementation details. Section 4 verifies the effectiveness of the proposed algorithm through experiments and compares it with the current mainstream algorithm. Finally, the thesis is summarized in section 5, and further research direction is pointed out.

## 2 RALATED WORK

The core problem of bas-relief generation is how to compress the 3d model while maintaining its original hierarchical relationship and detailed features. The related algorithms can be divided into two types: algorithms based on the depth domain and algorithms based on the gradient domain [6].

### 2.1 Bas-relief Generation Based on Depth Field

Algorithms based on depth field transform the height of the model with linear or nonlinear functions to map the original data to the base surface. Among them, Cognoni et al. [1] first studied this problem. They used a z-buffer algorithm in perspective projection to obtain the height values of bas-relief vertices. The implementation of the algorithm is simple and efficient, but it is easy to lose details when the compression ratio is large, and the effect for a complex model is not good. The following research considers the geometric feature while compressing the model height. For example, in [8], the adaptive histogram equalization (AHE) method in image processing was extended to the 3d model, and the compression was generated by gradient-weighted histogram equalization of the model depth value. The bas-reliefs generated by this method have good effects in feature maintenance and detail clarity. However, the time complexity of AHE is high; the algorithm needs to set six parameters, and the adjustment of any parameter requires recalculation of AHE. Therefore, it is less efficient in interactive adjustment. Li et al. [9] first decomposed the mesh surface in the frequency domain and then processed the mesh of different frequencies separately. The algorithm could retain details well, but the processing efficiency for a large-scale mesh surface was also not high. Zhang et al. [10] first extracted detailed features from the overall model via a bilateral filtering algorithm and then took these detailed features into special consideration. Because there is no need to solve complex differential equations, the algorithm has high execution efficiency, but the gradient information is not considered. Therefore, the bas-relief saturation is insufficient. Overall, the depth domain method does not involve conversion with the gradient domain, so the algorithm is fast and efficient. However, it only operates in the depth domain, which tends to lose the hierarchical relationship of the original model, resulting in poor saturation of the generated bas-relief model.

### 2.2 Bas-relief Generation Based on Gradient Field

This type of algorithm transforms the depth information of the model to the gradient domain, then processes it, and finally transforms the bas-relief model back to the depth domain through Poisson reconstruction. Among them, Song et al. [11] used Mesh Saliency [12] to measure the significance of visual features of the model and then used an Unsharp Masking operator to enhance the details. The Mesh Saliency is the description of the curvature change, so the algorithm can retain the detail features well. However, the algorithm assumes that the first partial derivative of the model is zero, which leads to obvious deformation in some parts of the result. Weyrich et al. [13] extended the gradient domain compression technology for high-dynamic-range images [14] and realized bas-relief production through nonlinear compression of the gradient amplitude. In this method, the compression amount of a region with a large gradient value is larger, while the compression amount of a region with a small gradient value is smaller, so the detail information can be retained. The authors also used boundary preserved diffusion filtering to deal with discontinuity in the gradient domain and used a multiscale method to distinguish the importance of different frequency features, which provides a large degree of freedom for bas-relief design. The bas-relief generated by the method has a good effect in contour definition, detail richness and other aspects, but the parameters used by the algorithm lack intuitive meaning and depend on the model, so more user-interaction settings are needed, and the overall efficiency is not high. Kerber et al. [15] applied the idea of literature [11] to the regular sampled gradient domain and used threshold values to filter out regions with excessive gradients. Because these regions usually appear at the contour and closed boundary, the above operation could make the transition part between different regions no longer occupy invalid depth space, thus reducing the depth change. The authors also used the Unsharp Masking to enhance the visual salience of small-scale features. On the whole, the method is simple and efficient, and the bas-reliefs generated have a good visual effect in terms of detail features. However, it is also prone to the problem that the features are excessively enhanced and the model is distorted. Subsequently, the authors improved the above method [16]. They used a bilateral filter to smooth the gradient. Because bilateral filtering is a boundary-preserving filter, its application can ensure the definition of the curvature extremum. This method significantly reduces the control parameters required by the system, and at the same time, it does not sacrifice too much bas-relief quality. To better restore the detailed information, Ji et al. [17] retained the local properties of the model by Laplacian operator. Wei et al. [18] first filtered the normal vector with Rolling Guidance based on a Gaussian Mixture Model and then reconstructed the bas-relief Model with the Surface from Gradients method. Christian et al. [19], Daniel et al. [20], and Zhang et al. [21] also considered the influence of illumination or observation perspective. In the previous work [22], we also provided a bas-relief generation algorithm with a point cloud as input, but the algorithm must be based on ordered grid-like control points, and saturation adjustment and real-time interaction cannot be realized. In addition to the above algorithms, many algorithms also pay special attention to execution efficiency and realize real-time operation with the help of GPU acceleration. For example, Kerber et al. [23] proposed two real-time adjustment algorithms for the depth domain and gradient domain. Zhou et al. [24] used a new nonlinear compression function to compress the gradient information. Ji et al. [25] created bas-reliefs by combining high compression with detail characterization of gradient fields. Because that algorithm focuses on the execution efficiency, the bas-relief effect generated is still far from the ordinary algorithm. Meanwhile, to the best of our knowledge, there is no algorithm that can realize real-time interactive adjustment based on CPU calculation power.

## 3 OUR ALGORITHM

### 3.1 Algorithm Flow

The gradient-based bas-relief generation algorithms achieve height compression by modifying the gradient amplitude of the model. For mesh surfaces, the connection relation of neighborhood vertices is known, and the gradient can be discretized into a linear combination of neighborhood vertices. However, point clouds do not contain any topology information, which means that the linear correlation between the gradient and the vertex coordinates cannot be established, and the traditional algorithm is no longer effective. Therefore, this paper takes the normal vector as the compression object, establishes the linear constraint relation between bas-relief vertices through the vertical relation between the local points and the normal vector, and then obtains the closed-form solution of the bas-relief model. The specific algorithm is shown in Fig. 2. Given a point cloud with normal vectors, the algorithm first finds the visible point cloud from the current perspective and calculates its curvature. Then, sparse control points are extracted from the visible point cloud, and the general shape of the bas-relief is outlined using the normal space compression strategy. Finally, the height mapping from control points to visible point cloud is realized through parametric surface fitting, and the bas-relief is adjusted to the desired height by an iterative algorithm. In the process, the algorithm provides parameters for interactive adjustment of bas-relief height, saturation and detail richness and achieves real-time adjustment through parallel processing.

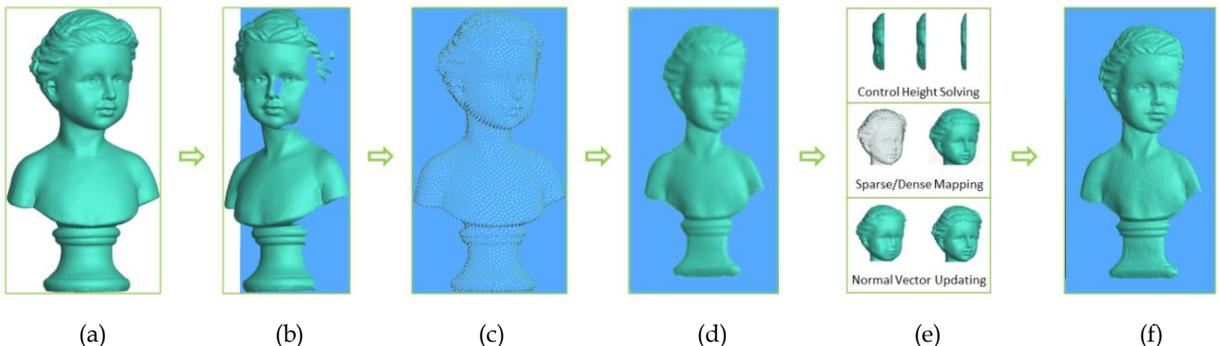

Fig. 2 Algorithm flow of this paper. Given a point cloud with normal vectors (a), our algorithm first detects the visible points and calculates its curvature (b). Then, sparse control points are uniformly sampled from the visible points (c), and an initial bas-relief can be reconstruct from them (d). In the following steps, the user can adjust the bas-relief effect in real-time (e), and finally, the algorithm can generate a bas-relief model with the desired height (f).

## 3.2 Visible Point Cloud and its Curvature

As in reference [22], we first use the Euler rotation shown in equation (1) to adjust the attitude of the entire point cloud so that the direction of the visual axis is parallel to the Z-axis of the coordinate system. In equation (1), $R_x$ and $R_y$ represent the orthogonal rotation matrix around the x and y-axes, $\theta_x$ is the angle between the perspective direction $v_r$ and the XOZ plane, and $\theta_y$ is the angle between $v_r$ and the Z-axis after the $R_x$ rotation. The above transformation makes it only necessary to consider the adjustment of z coordinate in the subsequent processing, which greatly reduces the computational complexity.

$$\mathbf{P}' = \mathbf{R_y}(\theta_y) \cdot \mathbf{R_x}(\theta_x) \cdot \mathbf{P} \qquad (1)$$

After completing the attitude adjustment, we divide the grid in the XOY plane with the side length of $2\rho$ ($\rho$ denotes the average sampling density of the point cloud) and project each point in the visible point cloud into the corresponding grid according to its x and y coordinates. Finally, we identify the points whose distance to the maximum z coordinate in each grid is less than $2\rho$ as the visible points.

According to the definition, mean curvature is an effective measure of local surface changes. Meanwhile, relevant studies have shown that curvature [26] is closely related to visual attention. Therefore, curvature can be used as the basis for bas-relief compression. However, curvature is a function of the second derivative of a surface, and its calculation is easily affected by noise. Fortunately, the moving least squares surface [27] is a representation robust to noise, and its projection process can effectively filter out the influence of noise. For this reason, the method proposed by Yang et al. [28] is adopted to calculate the curvature of the visible point cloud in this paper. Specifically, for an arbitrary point $p$ in given data, a moving least square surface $g(x)$ as shown in equation (2) is fitted, and then the mean curvature of point $k_{mean}$ can be calculated with equation (3), where $n(x)$ is the Gaussian weighted average normal vector of neighbor points $Q$, $\theta$ is the Gaussian weight function, $\nabla g(x)$ is the gradient of $g(x)$, and $H(x)$ is the Hessian matrix of $g(x)$.

$$g(x) = n(x)^T \left( \frac{\partial e(y, n(x))}{\partial y}\bigg|_{y=x} \right) \qquad (2)$$

$$e(y, n(x)) = \sum_{q \in Q} \left( (y-q)^T n(x) \right)^2 \theta(y, q)$$

$$k_{mean} = \frac{\nabla g(x) \cdot H(g(x)) \cdot \nabla^T g(x) - \|\nabla g(x)\|^2 \cdot Trace(H)}{\|\nabla g(x)\|^3} \qquad (3)$$

## 3.3 Control Points and Normal Space Compression

To realize the subsequent adjustments in real-time, we first extract sparse control points from the visible point cloud using the uniform sampling algorithm in [29]; we then process the control points and map the results to the visible point cloud.

It is difficult to preserve the details and smooth transition of a highly discontinuous region by directly compressing the spatial coordinates. Therefore, this paper proposes a compression strategy combining depth space and normal space, as shown in equation (4). The objective equation consists of a position constraint term and a normal constraint term, where the position term is used to move the boundary point to the bas-relief base, while the normal constraint term is used to eliminate the height jump in the model and retain the saturation and detail as much as possible. In equation (4), $P_b$ and $P_c$ represent the boundary points and the control points, respectively, $\hat{p}$ is the boundary point to be solved, $\tilde{p}$ is the projection of the boundary point on the base surface, $\hat{n}$ and $\tilde{n}$ represent the result normal vector and expected normal vector, respectively, and the subscript $z$ represents that only the height component needs to be considered.

$$\arg\min \left( \sum_{p \in P_b} \left( \|\hat{p}_z - \tilde{p}_z\| \right)^2 + \sum_{p \in P_c} \left( \|\hat{n}_p - \tilde{n}_p\| \right)^2 \right) \qquad (4)$$

Setting $\tilde{n}_q$ in equation (4) as the original normal of the point cloud can maintain the expressiveness of the original model to the greatest extent. However, this also means that only bas-relief models of fixed height can be generated. To realize the arbitrary adjustment of bas-relief height, it is necessary to put forward a method to compress the original normal space. Research reveals that the perception of a spatial surface by the human eye is nonuniform and nonlinear, and the visual system is usually attracted by the region with large local surface variation, while paying less attention to the flat region in the model. As mentioned above, the mean curvature represents the average variation of the local surface, so it can be used as a criterion for bas-relief compression. Based on the above analysis, the normal space compression scheme as shown in equation (5) is proposed in this paper.

$$\tilde{n} = w_k w_b n + (1 - w_k w_b)[0, 0, 1]^T \qquad (5)$$

In equation (5), $w_k$ is a weight coefficient related to the mean curvature, and its specific form is shown in equation (6).

$$w_k(k_{mean}) = 1 - \exp\left( -\left( \frac{\|k_{mean}\|}{\alpha \delta} \right)^\beta \right) \qquad (6)$$

where $\delta$ is the standard deviation of the normalized mean curvature and $\alpha$ and $\beta$ are the adjustment parameters.

Analysis of equation (6) shows that the normal vector of a point with larger mean curvature is closer to the original value after compression, so the surface information loss is smaller; meanwhile, the normal vector of a point with smaller mean curvature is closer to the normal vector of the base surface, making less contribution to bas-relief height and losing more surface information. Experiments show that the above strategy can effectively keep the saliency and saturation of details in the model unchanged and ensure the overall effect of the bas-relief.

The other parameter $w_b$ in equation (5) is a boundary weight function, as shown in equation (7), which is used to further eliminate the height mutation at the junction between the base plane and the bas-relief model and realize the smooth transition between them.

$$w_b(dist) = 1 - \exp\left(-\left(\frac{dist}{2\rho}\right)^2\right) \quad (7)$$

Fig. 3 shows a typical compression weight variation. It can be seen that the parameter $\alpha$ can change the degree of weight decline. When $\alpha$ =0, there is no compression of the normal space. Therefore, the bas-relief can keep the original height as much as possible. With the increase of $\alpha$, the proportion of normal space compression becomes larger and larger, and the corresponding bas-relief height decreases and smaller. At the same time, parameter $\beta$ can adjust the weight curve with the center located at $k_{mean} = \alpha\delta$. When $\beta$ =0, the compression ratio of all vertices tends to be consistent, and the bas-relief saturation is the best. As $\beta$ increases, the compression ratio of small-curvature points keeps increasing, while the compression ratio of large-curvature points keeps decreasing, so the saturation of the bas-relief can be significantly changed. In view of the significant effect of parameters $\alpha$ and $\beta$ in adjusting bas-relief height and saturation, they are named the height coefficient and saturation coefficient, respectively.

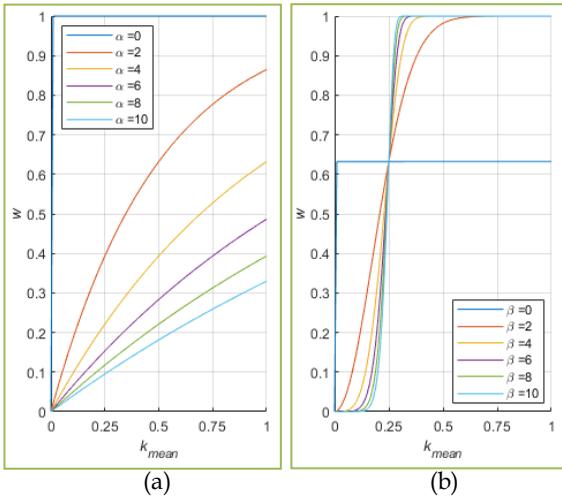

Fig. 3 Curvature weight variation with parameter $\alpha$ (a) and $\beta$ (b) when $\delta$ = 0.25

## 3.4 Solution of the Bas-relief Height

It is inefficient to directly optimize the objective equation (4). However, the control points and their neighborhood points can be approximately considered to be distributed in the same local plane. Therefore, the normal vector of the control points is always perpendicular to the spatial line formed by the current point and its neighbors. Furthermore, the equivalent description of the normal constraint term in equation (4) is that the expected normal vector should be as perpendicular as possible to the vector connecting the control point and its neighborhood points, that is, transformed into the form of equation (8):

$$\sum_{p \in P_c} \sum_{q \in Q} \left(\left\|\tilde{n}_p \cdot (\hat{p} - \hat{q})\right\|\right)^2 \quad (8)$$

where $Q$ represents the neighborhood point of the current point $p$. Finally, the objective equation shown in (9) can be obtained by the simultaneous equations (4) and (8).

$$\arg\min\left(\sum_{p \in P_b}\left(\|\hat{p}_z - \tilde{p}_z\|\right)^2 + \sum_{p \in P_c}\sum_{q \in Q}\left(\|\tilde{n}_p \cdot (\hat{p} - \hat{q})\|\right)^2\right) \quad (9)$$

In equation (9), each control point can be given at least k≥1 (in this paper, the number of neighborhood points is 6) normal constraint equations. At the same time, boundary points give more additional position constraints. Therefore, the number of equations is greater than the number of unknowns, belonging to an over determined equation. At the same time, when equation (9) is expanded, it is a linear combination of the height of control points, which can be arranged into the form AX = B, where A is a highly sparse matrix. Therefore, the closed-form solution can be obtained by solving the sparse equation [30] in the sense of least squares.

In the process of solving equation (9), the most time-consuming operation is triangular factorization of the semipositive definite matrix $A^T A$ [31], in which A is represented as the product of two triangular matrices $A^T A = R^T R$. Fortunately, for a specific bas-relief generation task, the matrix A is fixed, and its content does not change according to the parameter adjustment, namely, in the entire bas-relief generated process, the triangular factorization only needs to be performed one time. Once the decomposition is complete, the result can be used throughout the subsequent adjustment process. Therefore, the above decomposition can be implemented in advance before the formal adjustment, and the results can be retained to ensure the real-time solution of the control point's height.

## 3.5 Height Mapping and Detail Enhancement

After the control point heights are obtained, the results need to be mapped to the visible point cloud. Generally, there are two solutions. One solution is to fit the height of the control point by parametric surface and calculate the height of the visible point cloud on the parametric surface. The other solution is to first calculate the compression ratio of the control points and then map the compression ratio to the visible point cloud through the parametric surface. Because the control vertices are sparse, the height value of the parametric surface fitting cannot well predict the change of the unsampled part of the surface. In contrast, the compression ratio changes more gently, and there are two advantages to using this strategy. On the one hand, the fitting converges more easily; on the other hand, the detailed features between the feature points are easier to maintain through the proportional relation. The mapping results in Fig. 3 confirm the correctness of the above analysis. Therefore, the mapping scheme based on compression ratio is adopted in the following experiments.

In terms of the parametric surface used for the above mapping, there are many choices, such as distance weighting, radial basis function, and thin plate spline.

However, considering the demand of real-time processing, we choose multilevel–splines (MLSpline) [32]. Compared with other methods, MLSpline adopts an error refitting strategy, which gives it an extremely fast fitting speed. In addition, the fitting accuracy of MLSpline is only related to the number of fitting levels, so the accuracy of fitting results can be guaranteed. Later experiments show that MLSpline can completely meet the mapping requirements of this paper.

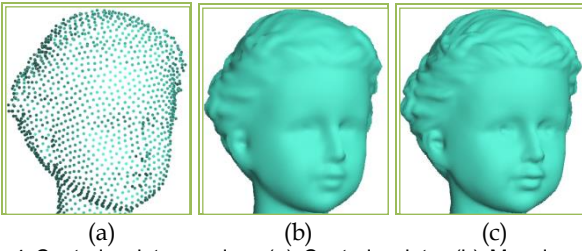

Fig. 4 Control point mapping. (a) Control points. (b) Mapping with height fitting. (c) Mapping with Compression ratio fitting.

Although the initial bas-relief from the control point mapping has good saturation, it is only a synoptic description of the original model, so further enhancement of the detailed features is necessary. Because the curvature has been calculated in advance, and the detail features are usually proportional to the variation of the local surface, we directly use equation (9) to calculate the increment of the height:

$$\Delta p_z = \gamma k_{mean} h \qquad (9)$$

where $\gamma$ is the detail coefficient and $h$ is the current bas-relief height. Subsequent experiments show that the detail amplitude of the bas-relief model is significantly enhanced after the above processing, and the model becomes more vivid and realistic.

It should be noted that the height mapping and detail enhancement can be performed simultaneously with the control point height solving, that is, in a continuous adjustment process, we can solve the height of control points at time t and simultaneously carry out height mapping and detail enhancement using the control points of t-1. This one-step-ahead control solution strategy allows the computation described in this section to run without requiring additional time on a multicore CPU. Another operation that can be performed in parallel is the updating of the bas-relief normal vector. Conventional normal vector calculation methods require neighborhood searching in 3d space and principal component analysis of the covariance matrix composed of neighborhood points, so they cannot meet the requirements of real-time processing. For this reason, we Delaunay the visible point cloud in the XOY plane [33][34] in advance and then update the normal vector of each triangle by calculating the cross product of its two edges. Finally, the normal vector of each vertex can be represented by the average normal vector of its 1-ring neighborhood triangles. Because the above method only needs to complete the low-complexity calculation, such as vector cross product and addition, it can be processed in real time.

## 3.6 Adjust bas-relief to a specific height

In some cases, such as combining multiple bas-reliefs into a whole, the height of the bas-relief needs to be controlled, that is, generate bas-reliefs that strictly conform to a specific height. From equations (6) and (9), it can be seen that the parameters affecting the bas-relief height include the height coefficient $\alpha$, saturation coefficient $\beta$ and detail coefficient $\gamma$. Among them, the effect of $\gamma$ is independent, that is, if the expected bas-relief height is $h$, then the bas-relief height only needs to be $(1-\gamma)h$ before detail enhancement. After removing the influence of $\gamma$, we can describe the problem as follows: making the bas-relief reach the specified height by adjusting the height coefficient while keeping the bas-relief saturation as much as possible. To this end, this paper will first set $\alpha$ to a very small value so that the initial bas-relief height is higher than the expected value; then, we increase $\alpha$ by a factor of two continuously. During the process, the bas-relief height gradually decreases and will fall below the desired height at last. At this point, we stop adjusting and record the current height coefficient $\alpha'$. Finally, in the interval $(\alpha'/2, \alpha')$, the binsearch method is used for iterative calculation until the height error is less than the set threshold. Because the bas-relief height is positively correlated with $\alpha$, the convergence of the above process can be guaranteed. At the same time, we added an outer loop to make the resulting model as saturated as possible by adjusting the value of $\beta$. The specific algorithm is shown in Algorithm 1.

**Algorithm 1: Adjust bas-relief to the desired height $h^0$**

**Input:** height coefficient $\alpha$, saturation coefficient $\beta$ and desired bas-relief height $h^0$

**OutPut:** height coefficient $\alpha$ and saturation coefficient $\beta$ that adujust bas-relief to the desired height $h^0$

1: Set: $\beta \leftarrow 0.00001$, $h_t \leftarrow \infty$
2: **while** ($||h_t - h^0||/h^0 > 0.01$)
3:  Set: $\alpha$, $\alpha' \leftarrow 0.001$, $\triangle h$, $h_t \leftarrow \infty$, $bFind \leftarrow$ *false*
4:  **while** ($\triangle h > 0.001$ and $|h_t - h^0|/h^0 > 0.01$)
5:   Solve bas-relief height by equation (8)
6:   Calculate current bas-relief height span $h_{t+1}$
7:   Set $\triangle h \leftarrow h_{t+1} - h_t$
8:   Set $h_t \leftarrow h_{t+1}$
9:   **if**(bFind == false)
10:    $bFind \leftarrow h_t < h^0$
11:   **if**(bFind)
12:    $\alpha' \leftarrow \alpha/2$
13:   **else**
14:    $\alpha \leftarrow 2\alpha$
15:   **end if**
16:  **else**
17:   $\alpha_{mid} \leftarrow (\alpha + \alpha')/2$
18:   **if**($h_t < h^0$)

19:     $\alpha \leftarrow (\alpha' + \alpha_{mid})/2$
20:     **else**
21:     $\alpha \leftarrow (\alpha + \alpha_{mid})/2$
22:     **end if**
23:   $\alpha' \leftarrow \alpha_{mid}$
24:   **end if**
25: **end while**
26: $\beta \leftarrow 2\beta$
27: **end while**

## 4 EXPERIMENT AND ANALYSIS

We implemented our algorithm with C++ and tested it on a PC with an Intel Core i7-8700 CPU and 16 GB of RAM. In the experiment, multithreading coding and OpenMP were used for parallel acceleration, and the CHOLMOD module in SuiteSparse was used to solve the objective equation (9). Additionally, unless otherwise specified, the values of each parameter are $\alpha$ =4.0, $\beta$ = 0.01, and $\gamma$ =0.02.

### 4.1 Elimination of High Discontinuity

One of the challenges in bas-relief generation is how to eliminate the high discontinuities in the visible point cloud. As shown in Fig. 5, although the depth domain method can treat this problem via operations such as histogram equalization and nonlinear transformation, the height mutation cannot be eliminated. Meanwhile, because the surface variation of the local surface is not considered, the bas-relief model is relatively flat and with poor expression of details. In contrast, the gradient domain method has a natural advantage in this respect. By maintaining the gradient continuity, the reconstructed model can realize a smooth transition on the jump region. Experiments show that our method can achieve the same effect. By searching neighborhood points in the 2d projection space, the algorithm can establish the connection between the point clouds on both sides of the jump region. Because the normal vector is the only constraint condition in this region, the high discontinuity can be eliminated by setting the expected normal vector on both sides to be the same. Furthermore, this characteristic can be passed to the visible point cloud via control point mapping. Compared with the gradient domain method, our method does not need to calculate the gradient information on the discrete point cloud, nor does it involve the depth/gradient domain conversion, so it is more intuitive and easier to implement.

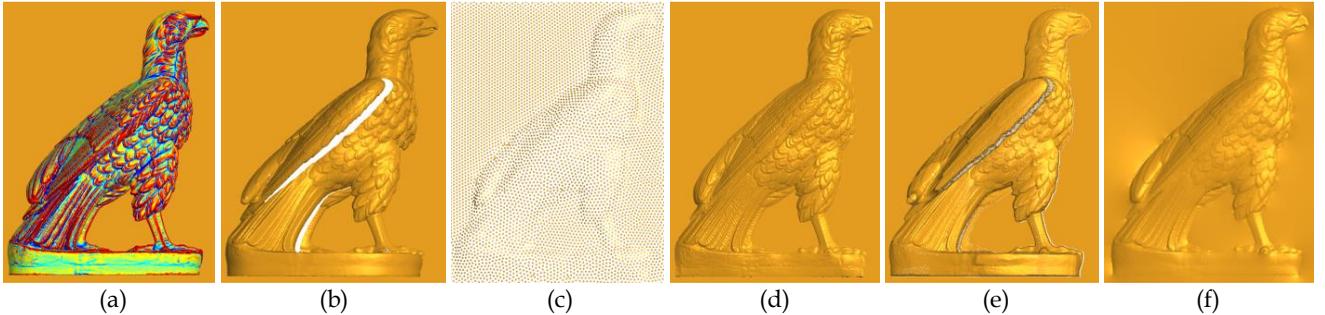

(a)    (b)    (c)    (d)    (e)    (f)
Fig. 5 Elimination of high discontinuities. (a) Original points. (b) Height discontinuity after visible point detection. (c) Control point result of our algorithm. (d) Bas-relief by our method. (e) Bas-relief by Global height histogram equalization [8]. (f) Bas-relief by Gradient domain compression [13].

### 4.2 Control point number and mapping method

As mentioned before, we used the method based on region growing in [29] to conduct control point sampling. Assume that the number of points before and after sampling is $n_1$ and $n_2$ and that the point cloud density is $r_1$ and $r_2$, respectively. Because the area of the sampling region remains unchanged before and after sampling, $n_1 r_1^2 = n_2 r_2^2$; that is, the region growing radius should be set as $r_2 = (n_1/n_2)^{1/2} r_1$. Although the number of control points obtained does not strictly meet the expectation, the above method has an efficiency advantage over the farthest sampling method. At the same time, the comparison of Fig. 6 (d) and (e) shows that our algorithm has a very weak requirement on the number of control points, and the variation of the number in a certain range will not significantly impact the bas-relief effect. In Fig. 6, the Gaussian weighted interpolation is also compared with MLSpline. The results show that MLSpline mapping can produce a smoother bas-relief effect. In addition, for the local shape of the model, such as the circular convex at the bottom of the model, the use of MLSpline can maintain the original shape to a greater extent. Moreover, the experiments of Fig. 6(a) ~ (d) show that, if the height discontinuity in the visible point cloud is removed using equation (9) in advance, the compression ratio variation between neighborhood points will be greatly reduced such that more ideal results can be obtained. Henceforth, this scheme is adopted in all of the subsequent experiments.

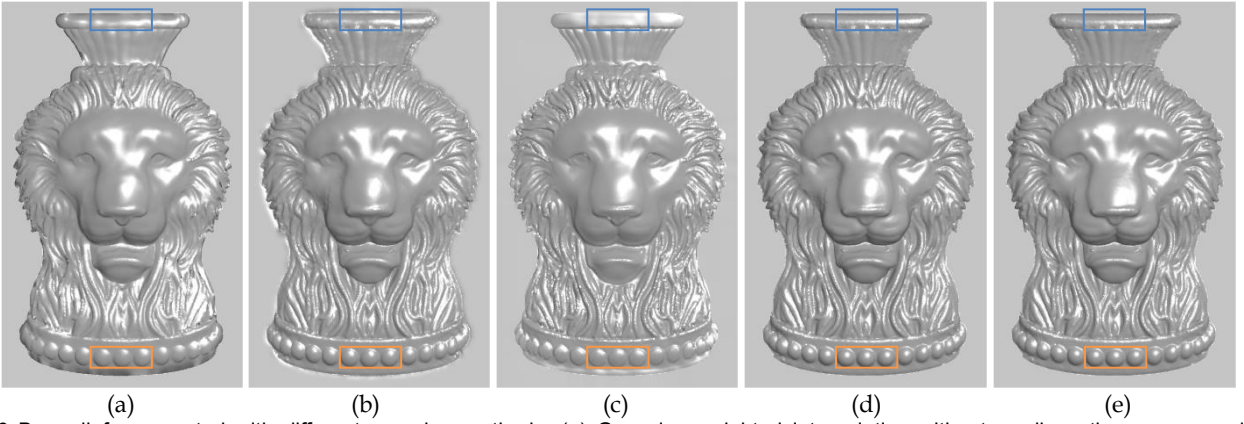

(a)     (b)     (c)     (d)     (e)

Fig. 6 Bas-reliefs generated with different mapping methods. (a) Gaussian weighted interpolation without prediscontinuous removal. (b) Gaussian weighted interpolation with prediscontinuous removal. (c) MLSpline mapping without prediscontinuous removal. (d) MLSpline mapping with prediscontinuous removal. (e) MLSpline mapping with half the number of control points in (d).

### 4.3 Bas-relief Adjustment

Fig. 7 shows the results of compressing the Armadillo model to different heights by adjusting the height coefficient. It can be seen from the results that the saturation and details of the bas-relief model at different heights are well maintained. Even in the extreme case that the height of the model is reduced to only $0.0125L_d$ (diagonal length of the bounding box), the detailed features, such as the feature lines of the head and lumps of the leg, are still clearly visible. Literature [25] also proposed an algorithm to enhance details through later processing. However, its initial bas-relief is based on linear compression in the depth field. Therefore, it is difficult to maintain the original saturation of the model under the condition of a large compression ratio. Especially for the model with few details in Fig. 8, the bas-relief results generated by the above algorithm are too flat in vision, thus losing the expressiveness of the original model. In contrast, the algorithm in this paper can maintain the height difference of different parts, thus generating more expressive bas-relief results.

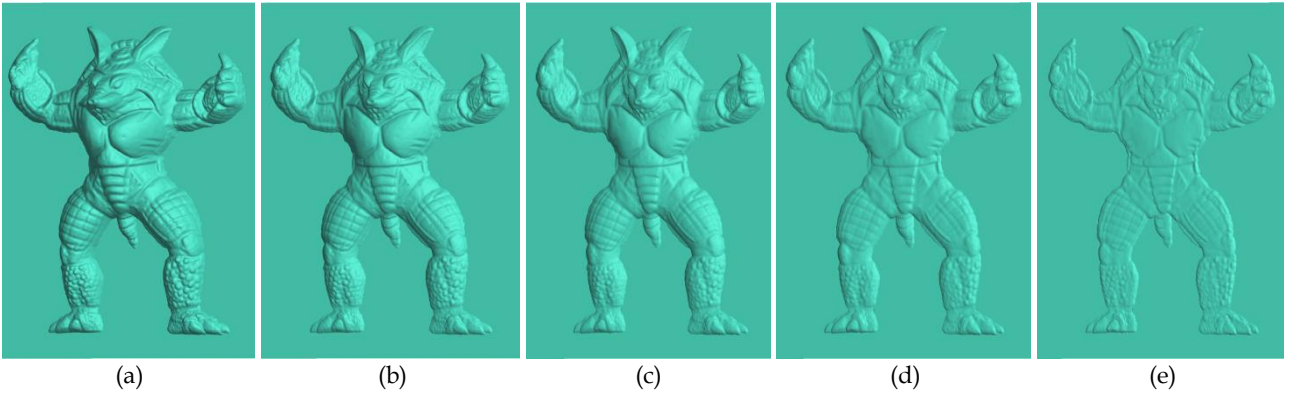

(a)     (b)     (c)     (d)     (e)

Fig. 7 Bas-reliefs of different heights. The heights of (a) ~ (e) are 20%, 10%, 5%, 2.5% and 1.25% of $L_d$, respectively.

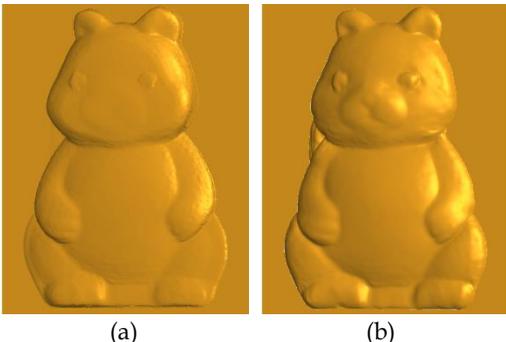

(a)         (b)

Fig. 8 Bas-relief generation with a height of $0.05L_d$ by algorithm [25] (a) and our method (b).

Fig. 9 shows a bas-relief model generated with different saturation coefficients. It can be seen that the smaller saturation coefficient has an advantage in maintaining the original saturation of the model, which allows the compression ratio of each point in the model to remain uniform, so that the height variation of each part in the generated model is relatively harmonious. With the increase of $\beta$, the height change of the flat region in the model decreases, while the height change of the high-curvature region becomes more intense, which makes the saturation of the model change significantly. For example, the horse's abdomen region is more protruding, while the neck becomes more flat, and the hip area of the model even gradually changes from a convex state to concave. The above changes indicate that parameter $\beta$ can effectively change the convex degree of the model such that the overall saturation changes significantly.

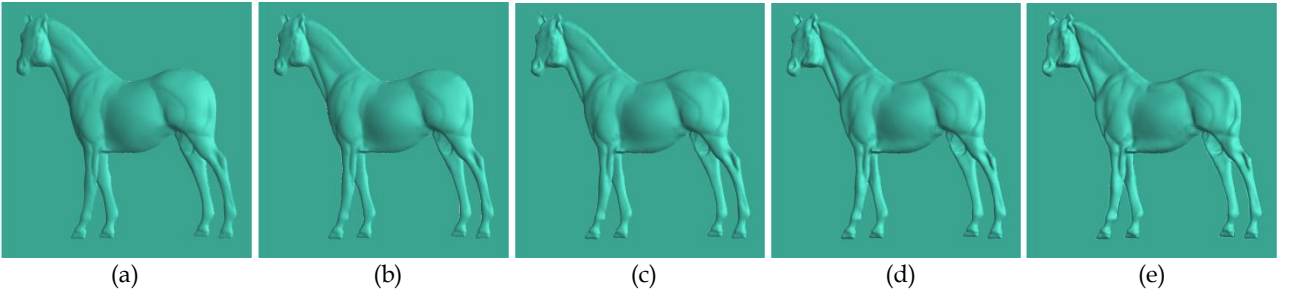

Fig. 9 Bas-reliefs generated with different degrees of saturation. (a) ~ (e) are bas-reliefs with $\beta$ =0, $\beta$ =2.5, $\beta$ =5.0, $\beta$ =7.5 and $\beta$ =10, respectively.

Fig. 10 shows the results of the detail enhancement of the bimba model. In the experiment, we created a constraint on the upper limit of the enhancement amplitude and suppressed the points with the top 1% curvature value to effectively overcome the deformation of the bas-relief model caused by the excessive enhancement amplitude of a few curvature significant points. Another solution for detail enhancement is to add constraints on the differential coordinates of the model to the target equation. However, this overall solution requires all visible points to participate to get a significant effect, so it cannot achieve real-time processing. In addition, it is also diffi- cult to accurately estimate the weight proportional to other constraint terms in the objective function. Comparatively speaking, the strategy of this paper is more flexible and can produce bas-relief results with more prominent details than those in the literature [13]. As analyzed above, although literature [25] also adopted a similar detailed enhancement strategy, its contour surface was generated based on linear compression and boundary smoothing filtering, so it could not well maintain the expressiveness of the original model, and the overall model was relatively flat.

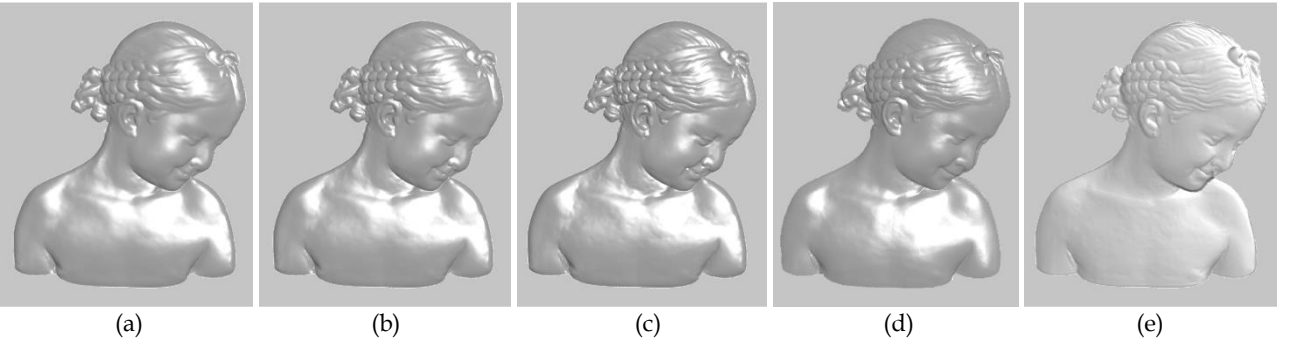

Fig. 10 Bas-reliefs generated with different degrees of detail. (a) ~ (c) are bas-reliefs with $\gamma$ =0.00, $\gamma$ =0.05 and $\gamma$ =0.10, respectively. (d), (e) are the bas-reliefs generated by [13] and [25] with feature reservation w = 0.5 (default value) and w = 10.

### 4.4 Curve-based Bas-relief

Compared with the bas-relief on the plane base, the bas-relief model created on the curved base is more varied and artistic. Therefore, it is popular in the field of art and decoration design. By replacing the boundary conditions in equation (4) with different forms, our algorithm can map the bas-relief model to different types of bases. In Fig. 11, we mapped the dragon model to the folded plane and the wave surface, respectively. It can be seen from the results that the bas-relief surface generated by the algorithm can closely follow the changes of the base surface, and the detailed information of the model is well preserved. In addition, our algorithm also supports the simultaneous projection of scene data containing multiple models onto the base surface. For example, in Fig. 12(a), we rotated the Armadillo model around its center of gravity by ± π /3, respectively, thus generating a point cloud with complex occlusion and overlapping relationships. Fig. 12(b) is a more challenging case. We put three independent models in the same scenario, so the depth variation range of the model is larger and the feature in- formation is richer. Nevertheless, our algorithm is still able to map the boundary regions of all models to the base surface accurately. Moreover, the depth jump region of a single model and the boundary regions of different models achieve a smooth and natural transition, resulting in a good, expressive bas-relief model.

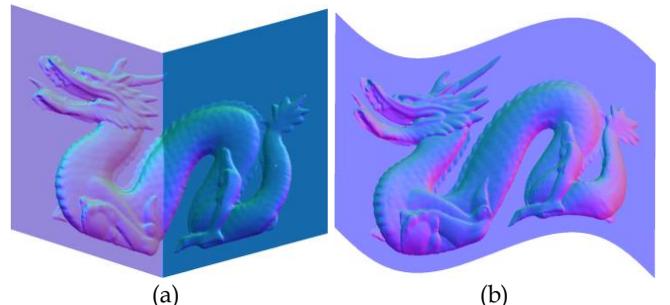

Fig. 11 Bas-relief result of different types of base. The dragon model is mapped to the folded plane (a) and the wave surface (b), respectively.

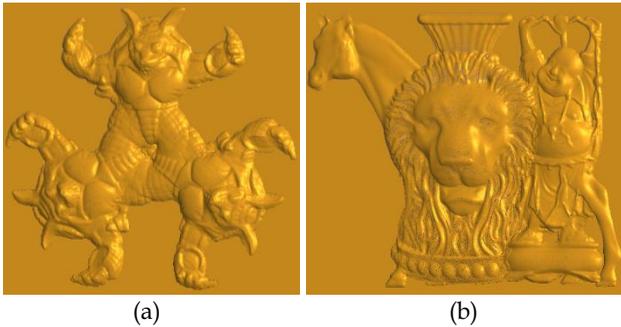

(a)           (b)

Fig. 12 Bas-relief models of a 3D scene containing multiple models.

More bas-relief models generated by our algorithm are shown in Fig. 13. It can be seen that all bas-relief models maintained good saturation and distinct levels, and the detailed features, such as hair and pattern, in the model have also been well maintained, which further verifies the effectiveness and universality of our algorithm.

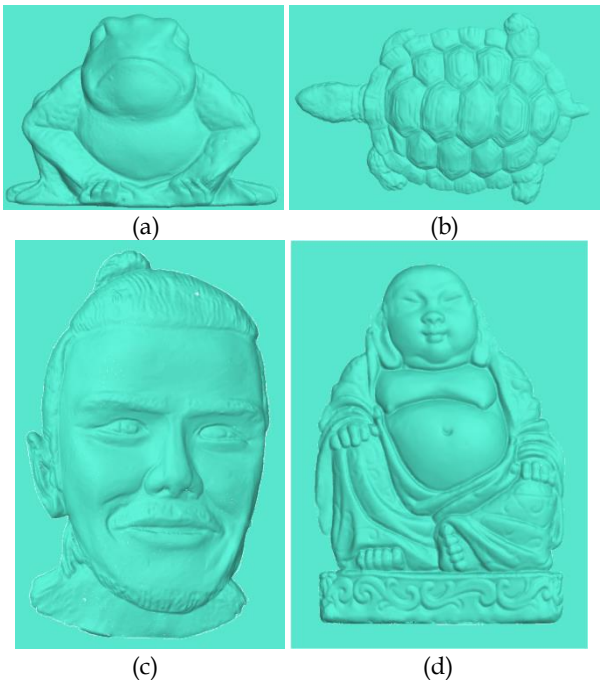

(a)           (b)

(c)           (d)

Fig. 13 More bas-relief models generated by our algorithm.

### 4.5 Computational efficiency

Table 1 lists the running time required by the algorithm in the interactive adjustment phase when processing different models. Because the three processes (solving the height of the control point, mapping the height of the visible point cloud, and updating the bas-relief normal) are carried out in parallel, the efficiency depends on the one that takes the longest time. Statistics show that the solution time is not sensitive to the change of the number of control points, and the time to solve the height of approximately 10,000 control points is approximately 20 ms, which can completely meet the speed requirements of real-time; for bas-relief models with a number of visible points not more than 340,000, the time for height mapping and normal update is controlled within 30 ms and, therefore, can also be processed in real-time. All this analysis reveals that the algorithm in this paper can achieve ideal results in the processing time and meet the requirements of real-time interactive modification in the process of bas-relief creation.

## 6 CONCLUSION

In this paper, a new bas-relief generation algorithm is proposed. The algorithm takes point cloud data as input and is suitable for the generation of different types of bas-reliefs. During the process, a small number of control points are used to first capture the outline of the bas-relief model, and then fine carving is used to enhance the saliency of the model details. Compared with solving the Poisson equation in the gradient domain, our algorithm directly compresses the normal space, thus overcoming the problem that the gradient cannot be represented as a linear combination of neighborhood coordinates in the discrete point cloud, making it more intuitive and easier to implement. The parameters in the algorithm have obvious physical significance and can flexibly modify the height, saturation and detail richness of the bas-relief. We also realized real-time adjustment, relying only on the computational power of a consumer CPU.

Currently, the algorithm only adopts a simple exponential transformation to adjust the bas-relief saturation. In the following work, we will design an interactive tool similar to brightness adjustment in the field of image processing to provide more diversified saturation adjustment schemes and a more creative space for users. At the same time, the algorithm only considers the curvature in normal space compression. In the next step, we will focus on the influence of lighting, color and other factors on the bas-relief effect and integrate them into the current bas-relief generation scheme. Finally, we will introduce more research results on a visual attention mechanism to further improve the visual effect of the bas-relief model.

### ACKNOWLEDGMENT

The authors would like to thank all the reviewers for their valuable comments. We also thank the AIM Shape Repository and the Stanford Repository for providing the models used in this paper. This work was supported in part by grants from National Nature Science Foundation of China (No. 61802204, 61571236, 61931012, 61602255).

TABLE 1
STATISTICS OF INTERACTIVE ADJUSTMENT TIME FOR EACH MODEL

| model | # visible pts | # control pts | t1(ms) | t2(ms) | t3(ms) | fps |
|---|---|---|---|---|---|---|
| children(Fig. 1) | 636699 | 7813 | 19 | 35 | 48 | 26.28 |
| bust (Fig. 2) | 104543 | 7924 | 17 | 26 | 7 | 38.38 |
| eagle (Fig. 5) | 388646 | 7913 | 18 | 33 | 27 | 30.16 |
| loin (Fig. 6) | 178120 | 6322 | 14 | 22 | 10 | 44.47 |
| armadillo (Fig. 7) | 320108 | 8129 | 20 | 32 | 25 | 31.70 |
| squirrel (Fig. 8) | 99331 | 6747 | 15 | 23 | 7 | 44.29 |
| horse (Fig. 9) | 184202 | 10045 | 23 | 35 | 18 | 28.51 |
| bimba (Fig. 10) | 198457 | 7469 | 16 | 26 | 13 | 36.39 |
| Dragon (Fig. 11) | 282472 | 8669 | 21 | 30 | 21 | 33.14 |
| models (Fig. 12.a) | 328846 | 10755 | 25 | 38 | 21 | 26.18 |
| models (Fig. 12.b) | 416327 | 9021 | 22 | 38 | 28 | 26.55 |
| frog (Fig. 13.a) | 168171 | 7664 | 17 | 27 | 12 | 36.66 |
| turtle (Fig. 13.b) | 181220 | 8111 | 17 | 28 | 12 | 35.34 |
| beckham(Fig. 13.c) | 346606 | 8089 | 18 | 33 | 24 | 30.33 |
| buddha (Fig. 13.d) | 173336 | 7314 | 17 | 26 | 11 | 38.35 |

Note: t1~t3 are times for control solving, height mapping, and normal updating, respectively.